# Development of a coronal mass ejection arrival time forecasting system using interplanetary scintillation observations

Kazumasa Iwai[*,1], Daikou Shiota[2,1], Munetoshi Tokumaru[1], Kenichi Fujiki[1], Mitsue Den[2], and Yûki Kubo[2]

1. Institute for Space-Earth Environmental Research, Nagoya University, Furo-cho, Chikusa-ku, Nagoya, 464-8601, Japan

2. National Institute of Information and Communications Technology, 4-2-1 Nukui-kita, Koganei, Tokyo 184-8795, Japan

[*] **Corresponding author: Kazumasa Iwai**

## Abstract

Coronal mass ejections (CMEs) cause disturbances in the environment of the Earth when they arrive at the Earth. However, the prediction of the arrival of CMEs still remains a challenge. We have developed an interplanetary scintillation (IPS) estimation system based on a global magnetohydrodynamic (MHD) simulation of the inner heliosphere to predict the arrival time of CMEs. In this system, the initial speed of a CME is roughly derived from white light coronagraph observations. Then, the propagation of the CME is calculated by a global MHD simulation. The IPS response is estimated by the three-dimensional density distribution of the inner heliosphere derived from the MHD simulation. The simulated IPS response is compared with the actual IPS observations made by the Institute for Space-Earth Environmental Research, Nagoya University, and shows good agreement with that observed. We demonstrated how the simulation system works using a halo CME event generated by a X9.3 flare observed on September 5, 2017. We find that the CME simulation that best estimates the IPS observation can more accurately predict the time of arrival of the CME at the Earth. These results suggest that the accuracy of the CME arrival time can be improved if our current MHD simulations include IPS data.

# Introduction

Solar wind and coronal mass ejections (CMEs) sometimes cause geospace disturbances when they arrive at the Earth. These disturbances are closely related to social infrastructures such as radio telecommunications, spacecraft and aircraft operation, and global positioning system (GPS)-based navigation. CME occurrences around the Sun have been observed using space-based white light coronagraphs (e.g., Yashiro et al. 2004). However, the initial speeds of CMEs derived from white light coronagraph observations are ambiguous because of line-of-sight projection effects. In addition, the acceleration and deceleration processes of CMEs propagating through interplanetary space are not well understood. These factors make the prediction of CME arrival difficult. There have been many models used to estimate the propagation of the CMEs. These include kinematic models and MHD simulations (Chen 1996, Gopalswamy et al. 2001, Vršnak and Gopalswamy 2002, Odstrcil et al 2003, Cargill, 2004, Shiota & Kataoka 2016). There are also some studies that predict the CME arrival time using space-based remote sensing and in situ observations of the inner heliosphere such as Solar TErrestrial RElations Observatory (STEREO) and MESSENGER spacecraft (e.g. Rollett et al. 2016; Moestl et al. 2017). Although the observations from these spacecrafts give important information about the propagation of CMEs in the inner heliosphere, their data are only available over limited time periods.

The solar wind includes a density disturbance, which causes the scattering of radio waves; this phenomenon is called interplanetary scintillation (IPS). Interplanetary CMEs (ICMEs) cause an increase in the amplitude of IPS because the sheath region, i.e., the high density and turbulent region in front of the CME, can significantly scatter radio emissions. Propagating CMEs have been observed by IPS observations (Tokumaru et al. 2000, 2003, Manoharan 2006, Iju et al. 2014, 2015, Glyantsev et al. 2015). In these previous studies, the CME propagation speed was derived from CME locations, which are determined from the radio source locations in the sky, and the time period that elapses between the CME occurrence near the solar surface and its IPS observation. Typical IPS telescopes scan the inner heliosphere once a day, making it difficult to detect the front of a fast-propagating CMEs more than one time. Only a few campaign-based observations have scanned the sky twice a day (e.g., Johri and Manoharan 2016) in order to determine CME locations at higher than a daily cadence.

The IPS amplitude has been estimated using solar wind models, and the solar wind itself has been re-constructed via the IPS tomography technique (Jackson et al. 1998, Kojima et al. 1998). CMEs are not prominent in the IPS tomography presented in these previous articles, which reconstruct the heliosphere assuming it corotates. However, a more recent time-dependent tomography analysis (Jackson et al 2003; Jackson et al. 2011 and references therein) can provide CME information, and is also used as an inner boundary for the three-dimensional (3D)-MHD simulations for forecasting CME structure (Yu et al. 2015), and arrival (Jackson et al 2015). Tokumaru et al. (2003, 2006) has also used a CME model to fit the IPS observations. This model-fitting technique requires only one day of IPS data, and it provides the 3D structure of CMEs in interplanetary space. The propagation of CMEs in interplanetary space is affected by the background solar wind. Therefore, a CME-solar wind model that integrates both the effects of background solar wind and CMEs (e.g., Chen 1996, Vršnak and Gopalswamy 2002) is important to accurately forecast their arrival at the Earth using IPS observations.

The global MHD simulations of the solar wind, which include the CMEs (e.g., Odstrcil 2003, Shiota and Kataoka 2016), enable us to estimate CME propagation and their interactions with background solar wind. These simulations provide the time variation of the electron density distributions of the inner heliosphere; knowledge of this variation enables us to estimate the scintillations of the radio emission.

The purpose of this study is to develop a system based on IPS analysis and global MHD simulations that can understand and refine the propagation processes of CMEs in the inner heliosphere and predict their arrival to the Earth more accurately that at the present. In our MHD simulation system, the propagation of the CME is calculated by the Space‐weather‐forecast‐Usable System Anchored by Numerical Operations and Observations (SUSANOO)-CME (Shiota and Kataoka 2016), and the IPS amplitude of each radio source is calculated using the 3D electron density variation derived from the MHD simulation. The estimated IPS response is compared with the observed IPS values to evaluate the accuracy of the MHD simulations.

## Method

**IPS observation**

The IPS observations are carried out by Institute for Space–Earth Environmental Research (ISEE), Nagoya University, Japan. ISEE operates radio telescopes dedicated to IPS observations at a radio frequency of 327 MHz. At this frequency, the IPS response under typical solar wind conditions can be approximated by weak scattering theory (Young, 1971) between 0.2 and 1.0 astronomical units (AU). IPS data obtained by the

Solar Wind Imaging Facility (SWIFT: Tokumaru et al. 2011) radio telescope of ISEE at the Toyokawa Observatory were employed in this study. This system has a fixed cylindrical parabolic reflector of dimensions 108 × 38 m, which has a single beam steerable between 60°S and 30°N along the local meridian, and observes 50 to 70 radio sources each day throughout the year.

The data processing of our IPS observations is described in reports of Tokumaru et al. (2000, 2003). We usually derive a scintillation level for each radio source which is the amplitude ratio of the scintillating component and noise component of the power spectra. Then, we derive the ratio of the instantaneous scintillation level to the typical scintillation level which corresponds to the g-value (Gapper et al. 1982) allowing the detection of the IPS enhancement caused by transient phenomena such as CMEs.

**MHD simulation (SUSANOO-CME)**

The MHD simulation in this study was originally developed by Shiota et al. (2014), and further expanded by Shiota and Kataoka (2016). This numerical code simulates the inner heliosphere between 25 and 425 solar radii (Rs) using a Yin-Yang grid (Kageyama and Sato 2004). The magnetic field at the inner boundary is derived from the potential field source surface (PSFF) model. The velocity, density, and temperature are derived from empirical models of the solar wind (Arge and Pizzo 2000, Hayashi et al. 2003). CMEs are included in the inner boundary of the simulation as spheromak-type magnetic flux ropes (SUSANOO-CME: Shiota and Kataoka 2016). The initial velocities of CMEs are roughly derived semiautomatically from the data of the Large Angle and Spectrometric Coronagraph (LASCO: Brueckner et al. 1995) onboard the Solar and Heliospheric Observatory (SOHO). For this study, several CME velocities within the error range are simulated to form an ensemble simulation set (Shiota et al. in prep). These simulations are evaluated using the IPS data, as described in the next section.

**IPS estimation using the MHD simulation**

We estimated the IPS *g*-value using the theoretical expressions of the radio scintillation assuming a weak scattering condition as given by Young (1971). The scintillation index (or so-called m-index) is the amplitude ratio of the total radio emission to the scintillating component of the radio emission. This index is obtained by integrating the density fluctuation of the solar wind ($\Delta N_e$) convolved with a weighting function of the IPS $w(z)$ along the line of sight to the radio source:

$$m^2 \propto \int_0^\infty \Delta N_e^2 w(z) dz, \qquad (1)$$

where z is the distance along the line of sight. The weighting function of the IPS is given by

$$w(z) = \int_0^\infty k^{1-q} \sin^2\left(\frac{k^2 z \lambda}{4\pi}\right) \exp\left(-\frac{k^2 z^2 \psi^2}{2}\right) dk, \qquad (2)$$

where $k$, $q$, $\lambda$, and $\psi$ are the wavenumber of density fluctuations in the solar wind, spectral index of the density disturbance, wavelength of the radio emission (0.92 m), and apparent size of the observed radio source. We set $q = 11/3$ because the density disturbance of the solar wind is shown to follow the Kolmogorov spectrum in the inner heliosphere (Woo and Armstrong, 1979). Following Tokumaru et al. (2003), we assumed all radio sources to be 0.1 arcsec in size. The disturbance level of the solar wind is assumed proportional to density ($\Delta N_e \propto N_e$). In addition, we assume the plasma density and election density have the same proportionality in our MHD simulation (i.e. electrical neutrality). Thus, we can estimate the scintillation index of radio sources from the total electron density along the line of sight by the following steps:

- Perform the MHD simulation as explained in the MHD simulation section but without the spheromak to obtain a time series of the 3D density distribution of the background solar wind.

- Estimate the scintillation index for a given radio source line of sight by integrating the convolution of the simulated electron density and the weighting function given by Equation 2, which we expect to correspond to the index from the background solar wind.

- Perform the MHD simulation with CMEs to obtain a time series of the density distribution of the inner heliosphere using SUSANOO-CME.

- Estimate the scintillation index using the density distribution with the weighting function along the line of sight from the combined background solar wind and CMEs.

- Normalize the simulated scintillation index from the solar wind with CMEs to that simulated from the solar wind without CMEs. This normalized scintillation index gives the IPS g-values in our MHD simulation.

We compare the simulated IPS g-values with the observed IPS g-values, and discuss this comparison in the following sections.

In the proposed system, we perform several simulations with different CME velocities and estimated IPS distributions from each simulation. Finally, the CME simulation that best reconstructs the IPS observation is used to forecast the CME arrival time.

## Results

We demonstrate our IPS-MHD simulation using a halo CME event generated by a X9.3 flare observed on September 5, 2017 (e.g., Shen et al 2018). The detailed analysis of the IPS data during this period is given by Tokumaru et al. (in prep). In this paper, we present the observed data and the derivation of the CME velocity from the viewpoint of space weather forecasting.

**IPS Observations**

The left panel of Figure 1 shows a difference image from the SOHO/LASCO white light coronagraph on September 6, 2017. The halo CME associated with the X9.3 flare is observed most prominently towards the south and west side of the Sun. The right panel in Figure 1 shows an all-sky map of the IPS *g*-value observed by the SWIFT radio telescope. In this figure, the location of each radio source is shown at a distance from the Sun scaled to the closest point on the line of sight to the radio source (the so called P-point). Hence, distance $D$ is expressed as $D = 1AU \times \sin\theta$, where $\theta$ is the elongation of the radio source. The center of the figure is the Sun, and the largest circle indicates 1 AU (the location of the Earth). The meridian scan observation of SWIFT started around 18 UT on September 6 from the west edge (right side) of the figure and ended around 10 UT on September 7 at the east edge (left side). Red, green, and blue diamonds indicate the radio sources for which the *g*-value is larger than 2.0, between 1.5 and 2.0, and between 1.2 and 1.5, respectively. The red and green diamonds on the southeast region indicate the location of the CME.

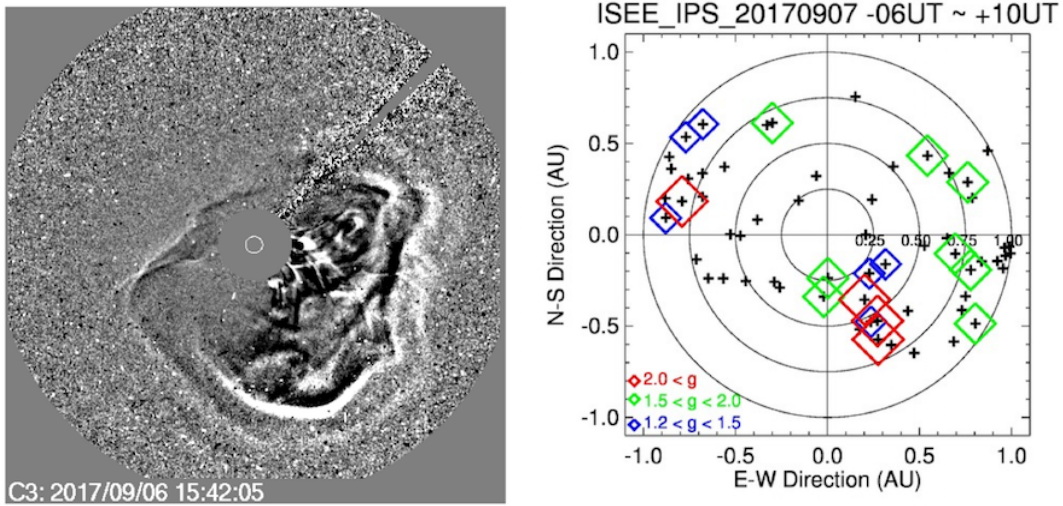

Figure 1

(title)

An example of the white light coronagraph image and all-sky map of the IPS g-value.

(legends)

(Left) An example of the white light coronagraph image observed by SOHO/LASCO on September 6, 2017. (Right) An all-sky map of the IPS g-value observed by SWIFT on September 7, 2017. The symbols indicate +: all observed radio sources, diamonds: radio sources with *g*-values as follows: $2.0 < g$ (red), $1.5 < g < 2.0$ (green), and $1.2 < g < 1.5$ (blue).

**MHD simulation and IPS estimation**

Figure 2a shows the scintillation index of the background solar wind estimated by SUSANOO. Figure 2b shows the estimated scintillation index by SUSANOO that includes the CMEs that occurred in early September 2017. The ratio of the simulated indices shown in Figure 2b relative to those shown in Figure 2a gives an estimation of the *g*-values, which are plotted in Figure 2c. The distribution of the estimated g-values are compared with those of the observed *g*-values (Figure 2d) using the same coordinate system (apparent elongation from the Sun in radians). We assumed the size of all radio sources are 0.1 arcsec to calculate the scintillation level, the usual value ascribed to most scintillating sources observed at radio frequencies of 327 MHz (e.g. Tokumaru et al. 2003). Although the estimated scintillation index may be different from the actual one because of this, the estimated g-value in Figure 2c does not exhibit the source size effect because the ratio of the scintillation indices calculated with and without CMEs cancels this effect. Note that we assume each spheromak is directly related to the position of the flare onset where the spheromak extends outward radially. These assumptions may limit the accuracy of the predicted g-value distributions, because of the fact that the flare location itself is often not the location of the ejected material, and non-radial flow in the corona is often known to be present. The difference between the spheromak and the actual CME shape also causes the different g-value distributions. Although the detailed analysis of the CME deflection is beyond the scope of this study, it should be mentioned that Shiota and Kataoka (2016) pointed out the deflection of the spheromak in the MHD simulation can affect the CME predictions.

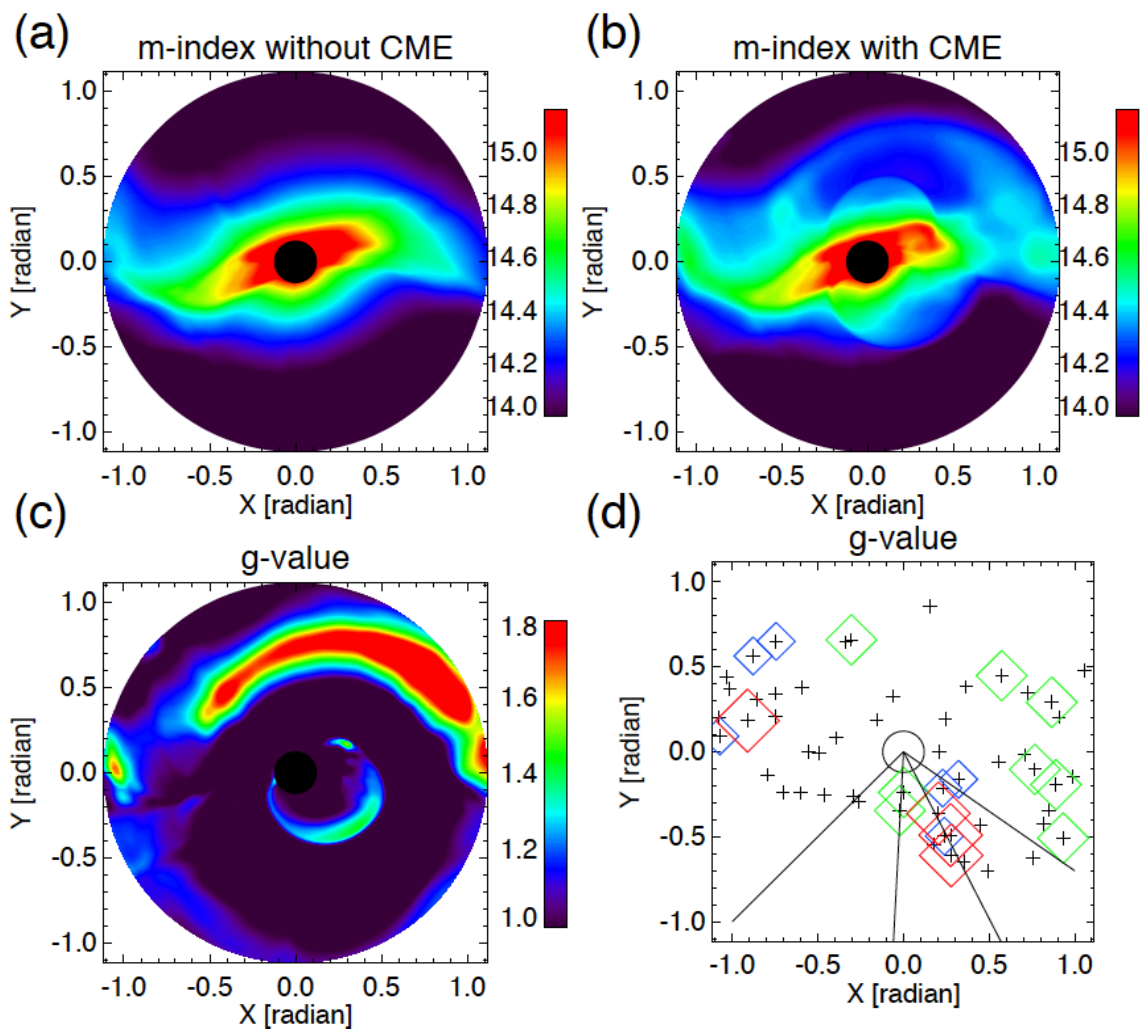

Figure 2

(title)

Scintillation index and g-values estimated by SUSANOO and observed by SWIFT

(legends)

(a) Scintillation index (Log scale) of the background solar wind without CMEs estimated by SUSANOO at 1:00 UT on September 7, 2017. (b) Scintillation index (Log scale) of the solar wind with CMEs estimated by SUSANOO-CME. (c) Estimated g-values, i.e., the ratio of the values shown in (b) to those shown in (a). (d) The IPS *g*-values observed by SWIFT displayed using the same coordinate system. Four solid lines indicate vertical axis of four panels in Figure 5. The vertical and horizontal axes in the figure indicate the elongation from the Sun (in radians).

**Parameter survey of the CME speed**

We estimate the initial conditions of the CMEs from SOHO/LASCO observations and include the corresponding spheromaks in the simulation. Two CMEs occurred in the period of interest—one was generated by a M5.5 flare on September 4 (the first CME) and the other, by a X9.3 flare on September 6 (the second CME). Table 1 summarizes the parameters of the two CMEs. Only the velocity of the second CME is a free parameter in this study.

Table 1 Parameters of the spheromak included as the CME in the MHD simulation

|  | Onset time | Velocity | longitude | Latitude | width 1 | width 2 | B (Mx)[1] |
|---|---|---|---|---|---|---|---|
| First CME | 0904 20:28 | 1000 km/s | 11° | 5° | 8° | 60° | 1.6e+21 |
| Second CME | 0906 11:53 | 1000 km/s ~ 2500 km/s | 23° | 7° | 3° | 80° | 3.0e+21 |

We performed several simulations for the second CME with different initial velocities ranging from 1000 km/s to 2500 km/s. In this paper, we consider four of these simulations: (RUN1) 1000 km/s, (RUN2) 1500 km/s, (RUN3) 2000 km/s, and (RUN4) 2500 km/s. Figure 3 shows the sky map of the IPS estimated from the four simulations. We are able to estimate the IPS for any time and any direction of interest because the global MHD simulation of SUSANOO provides the time variation of the 3D electron density between 25 and 425 Rs. On the other hand, in an actual IPS observation by ISEE, the sky is scanned using the diurnal motion of the radio sources. Therefore, we only overplot the actual IPS observed within 1 h of each simulated time.

---
[1] total magnetic flux contained in the spheromak (Mx)

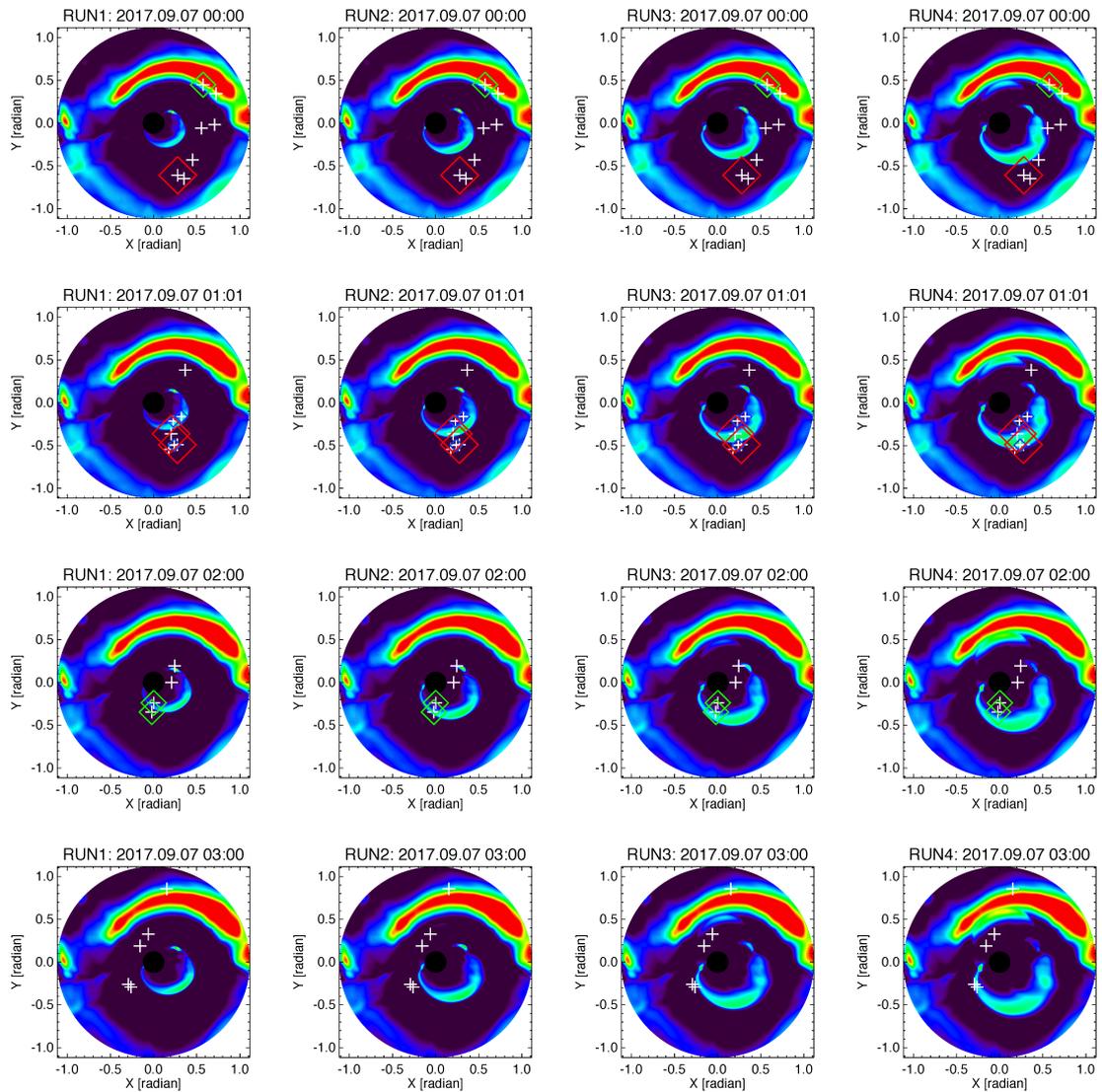

Figure 3

(title) Time variation of the SUSANOO-estimated g-value (the background) and observed IPS *g*-values (symbols)

(legends) Time variation of the SUSANOO-estimated (the background) and observed

(symbols) IPS *g*-values on September 9, 2017. Simulations (columns from the left): RUN1, RUN2, RUN3, and RUN4. Time (rows from the top: 0:00 UT, 1:00 UT, 2:00 UT, and 3:00 UT.

## Discussion

**Physical meaning of the observed and estimated IPS values**

The observed *g*-value is derived by fitting the scintillation index over the source elongation. First, we derive a typical scintillation index curve at given elongation angle for each radio source by fitting the daily variation of the scintillation index using $\bar{m} = aD^{-b}$, where $a$ and $b$ are constants, and $D$ is the distance between the Sun and the closest point of the line of sight of the radio source. The ratio of $m$ and $\bar{m}$ at a given elongation angle is defined as the *g*-value. The enhancement of the observed *g*-value indicates a density enhancement along the line of sight by phenomena such as CMEs, heliosphereic current sheets (HCS), and corotating interaction regions (CIR). On the other hand, the *g*-value estimated from our simulation is the ratio of the scintillation index with and without CMEs at the same time and direction. Therefore, our simulated g-values cannot be used to detect phenomena such as HCSs and CIRs that last longer than 1 day; this estimation is suitable for extracting the g-value enhancement generated only by

CMEs. In our comparisons, the errors induced by the unavoidable inclusion of phenomena such as HCSs and CIRs in the observed g-values are assumed small.

**Spatial variation of the estimated *g*-value**

In our MHD simulation, the g-value enhancement shows a loop-like or toroidal distribution in agreement with the observed IPS. Figure 4 shows the density distribution on the ecliptic plane estimated by RUN2 on September 7 1:00 UT (same time as that depicted in Figure 2). In this analysis the enhanced g-value region is present at the front of the propagating spheromak where the density is higher than that of the surrounding solar wind. The initial spheromak does not contain any density enhancement (i.e., the density of the spheromak is equal to that of the background solar wind). The density enhancement in our simulation is generated by the compression of the solar wind swept by the higher-speed CMEs. This is consistent with the explanation given by Tokumaru et al. 2003 (see Figure 4 in Tokumaru et al. 2003).

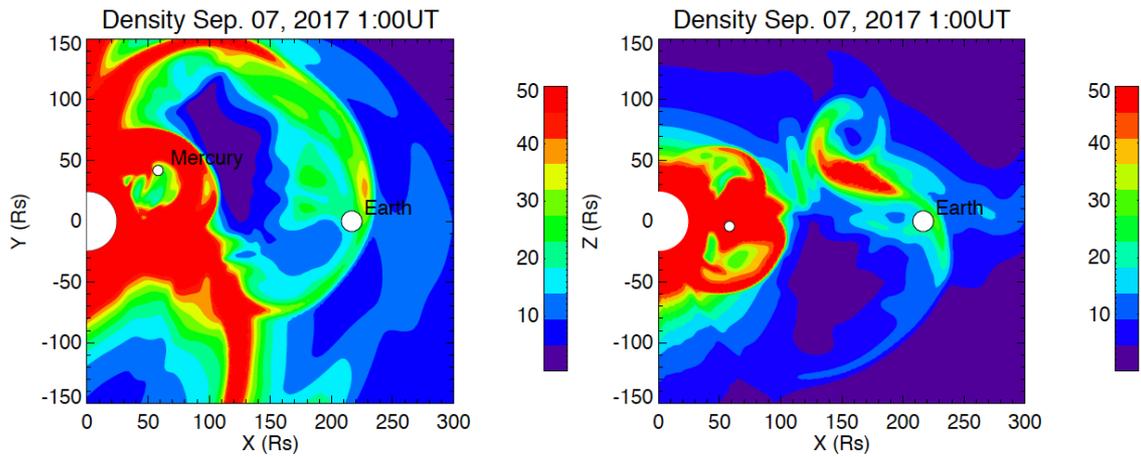

Figure 4

(title)

Density distribution on the ecliptic plane (left) and Sun-Earth plane (right) estimated by RUN2

(legends)

(Left) density distribution on the ecliptic plane estimated by RUN2 at 1:00 UT. (Right) density distribution on the Sun-Earth plane inclined 90° from the ecliptic plane. Location of the Earth is at [216, 0] as indicated by the large white circle. A smaller white circle estimates the location of Mercury projected onto these planes. Horizontal and vertical axes: distance from the Sun (Rs). Color: electron density (/cc).

Figure 3 shows two groups of *g*-value enhancements—one in the northern hemisphere caused from the first CME, and the other in the southern hemisphere caused from the second CME. Although the simulation in Figure 4 shows that the first CME had already arrived at the Earth, a "backside lobe" of the spheromak still remains near the Earth at this time and caused the g-value enhancement. If we adopt the traditional assumption that the CME should be at the P-point, which is the closest point along the line of sight of the radio source, the first CME should be located between 0.5–0.8 AU at the time depicted in Figure 4. Without this figure, we might mistake that the high g-values in the northern hemisphere are caused by the second CME. Our MHD-based IPS simulation shows that we can clearly distinguish the *g*-value enhancements originating from the first and second CMEs. Previously, it has been difficult to distinguish the IPS enhancements generated by multiple CMEs using all-sky maps with 1-day cadence. The IPS simulation developed in this study can be applied to studies of multiple CME events, such as CME-CME interactions, in future studies.

**Best CME velocity**

As shown in the previous sections, the simulated and observed *g*-values were derived using different approximations. This may present differences in the absolute *g*-values. Therefore, in further certification of our simulations, we use the location of the shock front of the CME to fit the CME speed. Figure 5 shows the relationship between the observed and estimated *g*-values along a specific declination angle from the solar north pole. At 1:00 UT, two high *g*-value sources (red lines in Figure 5a) are observed. The shock of RUNs 2, 3, and 4 crossed at least 1 radio source with a high g-value. Weak *g*-value sources that correspond to the backside of the shock (blue lines in Figure 5b) are also observed. At 2:00 UT, the shock of RUN 3 and 4 has already passed the high *g*-value sources (green lines in Figure 5c), while that of RUNs 1 and 2 are just crossing these sources. At 3:00 UT, the shock of RUN 4 is just crossing a radio source without a high *g*-value (black line in Figure 5d). It seems that RUN2 or RUN3 fit the observed g-values best although both of them are not the perfect fits.

In order to provide a more objective determination of the best fit to the observed g-value, we use simple indices corresponding to the number of observed radio sources whose *g*-value is consistently estimated by MHD simulation. We define that the g-value is consistently estimated such that both estimated and observed *g*-values are larger than 1 (true positive: TP), or both of them are smaller than or equal to 1 (true negative: TN). The accuracy of the fitting is defined as the ratio between the number of consistently estimated radio source (TP + TN) and total number of the observed radio sources around the CME. We also use a skill score that finds if g-value is inconsistently estimated; the MHD estimates g>1 while the observed g-value was g≤1 (false positive: FP), and the MHD estimated g≤1, while the observed g-value was g>1 (false negative: FN). The true skill statistic (TSS) is defined, TSS=TP/(TP+FN)-FP/(FP+TN). We also evaluate the average of the positional difference between the g-value peak and radio sources with g>1 along the same position angle. Table 2 summarizes the indices of the four RUNs. RUN2 provides the best match among them.

Table 2 Initial speed (second CME) and the observed radio sources with consistent *g*-values from MHD simulation.

|  | RUN1 | RUN2 | RUN3 | RUN4 |
|---|---|---|---|---|
| CME speed | 1000 km/s | 1500 km/s | 2000 km/s | 2500 km/s |
| Accuracy | 0.6 | 0.8 | 0.6 | 0.5 |
| TSS | 0.4 | 0.7 | 0.4 | -0.3 |
| Average Difference (Radian) | 0.60 | 0.55 | 0.65 | 0.75 |

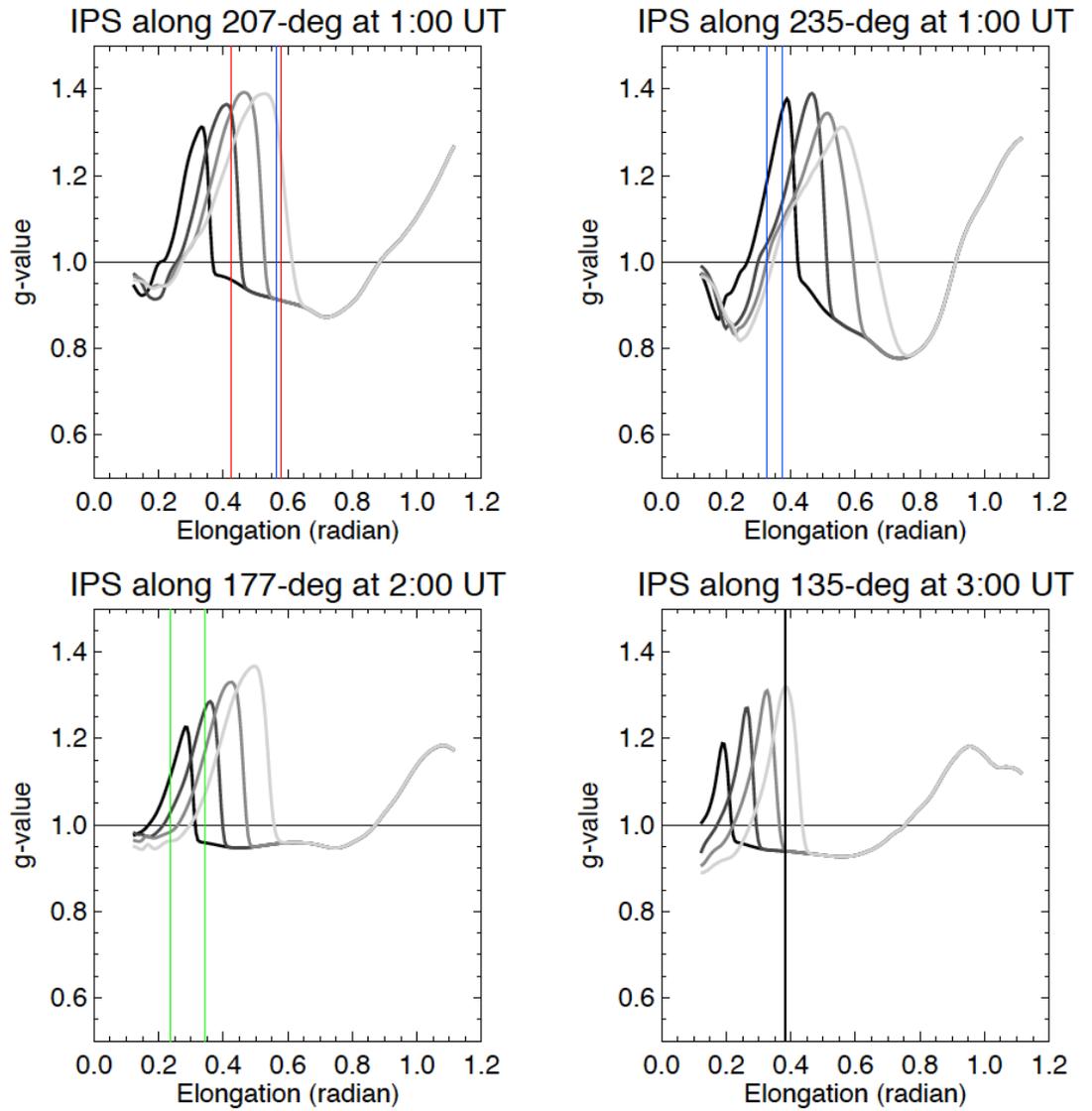

Figure 5

(title)

Slices of the estimated *g*-value along a specific position angle from the solar north pole.

(legends) Slices of the estimated *g*-value along a specific position angle from the solar north pole indicated by the solid lines in Figure 2d. (a) Along 207° at 1:00 UT,

(b) along 235° at 2:00 UT, (c) along 177° at 2:00 UT, and (d) along 135° at 3:00 UT. RUNs 1 to 4 are indicated by the black to gray lines. Vertical lines indicate the elongation of nearby radio sources along the slice observed at the same time: red: $2.0 < g$, green: $1.5 < g < 2.0$, blue: $1.2 < g < 1.5$, and black $g < 1.0$.

Figure 6 shows the time-distance plot of the velocity along the Sun-Earth line in RUNs 1–4. The two CMEs are recognized as two propagating highspeed regions. The arrival time of the second CME is defined by the shock arrival time observed by the Deep Space Climate ObseRvatory (DSCOVR) spacecraft, an in situ instrument at the Lagrange point (L1: 213 Rs), which is indicated by the horizontal line in Figure 6. The arrival time of the CME in RUNs 1, 2, 3, and 4 are around 24:00 UT, 22:00 UT, 18:00 TU, and 14:00 UT on September 7, respectively. The actual CME shock arrived at the Earth around 22:28 UT on September 7, 2017 (Shen et al. 2018), and is indicated by the vertical line in Figure 6. Therefore, RUN2 is the best fit and this result is consistent with the best fit of the estimated IPS as shown in Figure 5 and Table 2. In the CME list automatically generated by CACTus (Robbrecht et al. 2009), the median velocity and highest velocity of the second CME are 978 and 1955 km/s, respectively. These CME velocities correspond to RUN1 and RUN3, respectively. Therefore, our results suggest that the initial speed of the

CME automatically derived by the white light coronagraph images are validated by the IPS data.

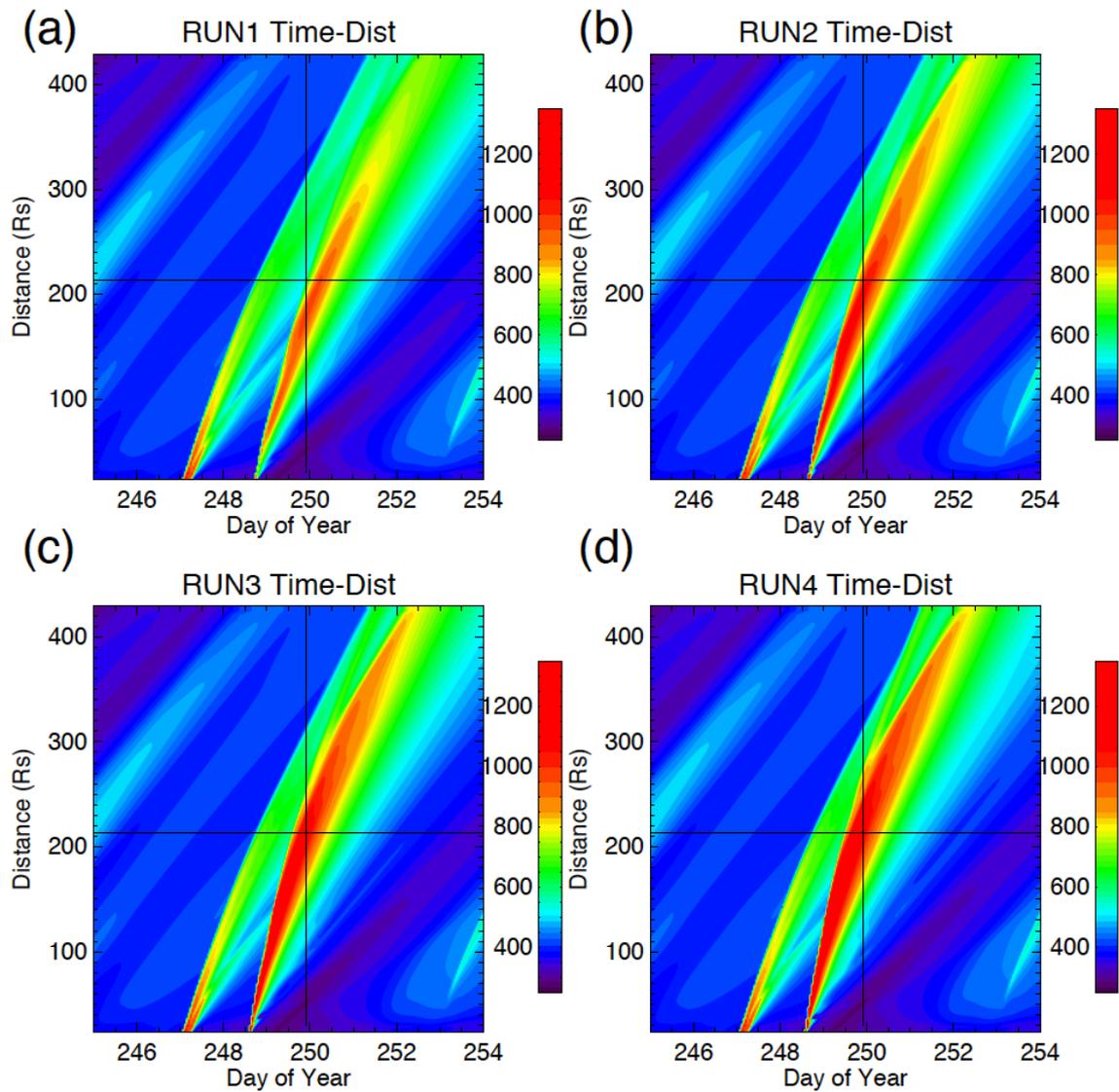

Figure 6

(title)

Time-distance plot of the velocity derived from the SUSANOO-CME along the

Sun–Earth line.

(legends) Time-distance plot of the velocity derived from the SUSANOO-CME along the Sun–Earth line. Vertical axis: distance from the Sun (RS). Horizontal axis: Day of year in 2017. Color contour: plasma velocity (km/s). (a) RUN1, (b), RUN2, (c) RUN3, and (d) RUN4. The horizontal line at 216 Rs indicates the location of the Earth, and the vertical line at 22:30 UT on September 7, 2017 indicates the shock arrival time.

**Concluding remarks**

We have developed an IPS estimation system based on the global MHD simulation SUSANOO-CME. The scintillation index was estimated using the density distribution in the inner heliosphere derived by the MHD simulation. Then, the *g*-value enhancement by CMEs was estimated by using the ratio of the scintillation index with and without CMEs. The simulated IPS values are compared with the IPS observations made by ISEE, Nagoya University. The simulated *g*-values show good agreement with the observed *g*-value. We simulated several *g*-value time variations generated corresponding to different initial speeds of the CME. We found that the CME that shows the best fit to the IPS observations allows a forecast of the arrival time of the CME at the Earth most accurately. Although a quantitative assessment of the accuracy of this system is beyond the scope of this study, our result suggests that the accuracy of the CME arrival time could be improved if the current MHD simulations include IPS data.

From the viewpoint of space weather forecasting, stable real-time predictions are important. Our MHD simulation system based on the IPS observation has a large advantage to the stable operation of the forecasting system because the ground-based IPS observations are more stable compared to the space-based observations. This system is included in the CME forecasting system developed by Shiota et al. (in prep). In their system, the initial condition of CMEs (parameters of the spheromak) is roughly derived semiautomatically from the space-based coronagraph observations and the location of the associated flare on the solar surface. Then, many CMEs with possible parameter sets are simulated suing SUSANOO. Each simulation result is evaluated by the IPS observations using our IPS simulation system. This real-time forecasting system is expected to improve significantly the accuracy of the arrival time of the CME to the Earth.

# Declarations

**Ethics approval and consent to participate**

Not applicable

**Consent for publication**

Not applicable

**List of abbreviations**

CIR: Corotating interaction region

CME: Coronal mass ejection

HCS: Heliosphereic current sheet

IPS: Interplanetary scintillation

LASCO: Large Angle and Spectrometric Coronagraph

MHD: Magnetohydrodynamic

SOHO: Solar and Heliospheric Observatory

SUSANOO: Space-weather-forecast-Usable System Anchored by Numerical Operations and Observations

SWIFT: Solar Wind Imaging Facility

**Availability of data and material**

The datasets supporting the conclusions of this article are available in the data repository of ISEE, Nagoya University http://stsw1.isee.nagoya-u.ac.jp/ips_data-e.html, and Virtual Solar Observatory https://sdac.virtualsolar.org/cgi/search.

**Competing interests**

The authors declare that they have no competing interests.

**Funding**

This work was supported by the Ministry of Education, Culture, Sports, Science and Technology (MEXT), Japan Society for the Promotion of Science (JSPS), KAKENHI Grant Number 18H04442, 15H05813 and 15H05814.

**Authors' contributions**

KI led this study and drafted the manuscript. DS developed the MHD simulation codes. KI, MT, and KF operated and maintained the IPS radio observations. DS, MD, and YK operated and maintained the space weather forecasting system based on the MHD simulation. All authors read and approved the final manuscript.

**Acknowledgements**

This work was supported by MEXT/JSPS KAKENHI Grant Number 18H04442, 15H05813 and 15H05814. The IPS observations were provided by the solar wind program of the Institute for Space-Earth Environmental Research (ISEE). This study was carried by using the computational resource of the Center for Integrated Data Science, ISEE, Nagoya University through the joint research program. We thank the LASCO coronagraph group for the white-light CME images.